\documentclass[twocolumn,preprintnumbers,superscriptaddress,nofootinbib]{revtex4-1}
\usepackage[utf8]{inputenc}
\usepackage{amsmath}
\usepackage{amsthm}
\usepackage{float}
\usepackage{braket}
\usepackage{graphicx}
\usepackage{amssymb}
\usepackage{url}
\usepackage[colorlinks=true, pdfstartview=FitV, linkcolor=red, citecolor=blue, urlcolor=blue]{hyperref}
\usepackage{slashed}
\usepackage[normalem]{ulem}

\newcommand{\dd}{\mathrm{d}}
\newcommand{\en}{\mathcal{E}}

\begin{document}

\title{Entropy Suppression through Quantum Interference in Electric Pulses}

\author{Gerald V. Dunne}
\email[]{gerald.dunne@uconn.edu}
\affiliation{Department of Physics, University of Connecticut, Storrs, CT 06269-3046, USA}

\author{Adrien Florio}
\email[]{aflorio@bnl.gov}
\affiliation{Center for Nuclear Theory, Department of Physics and Astronomy, Stony Brook University, Stony Brook, New York 11794-3800, USA}
\affiliation{Department of Physics, Brookhaven National Laboratory, Upton, New York 11973-5000}

\author{Dmitri E. Kharzeev}
\email[]{dmitri.kharzeev@stonybrook.edu}
\affiliation{Center for Nuclear Theory, Department of Physics and Astronomy, Stony Brook University, Stony Brook, New York 11794-3800, USA}
\affiliation{Department of Physics, Brookhaven National Laboratory, Upton, New York 11973-5000}



\bibliographystyle{h-physrev4}

\begin{abstract}
The Schwinger process in strong electric fields creates particles and antiparticles that are entangled.  The entropy of entanglement between particles and antiparticles has been found to be equal to the statistical Gibbs entropy of the produced system. Here we study the effect of quantum  interference in sequences of electric pulses, and show that quantum interference \textit{suppresses} the entanglement entropy of the created quantum state. This is potentially relevant to quantum-enhanced classical communications. Our results can be extended to a wide variety of two-level quantum systems.
\end{abstract}

\maketitle

\textit{Introduction.} Entanglement is the distinctive feature of the quantum world. It is also the core resource quantum computations rely upon to achieve quantum advantage. A key process to harvesting this resource is the generation of entangled states.

The question we address in this work is the following: what is the effect of quantum interference on the entanglement of particles created by a sequence of electric pulses? Specifically, we consider Schwinger pair creation in an electric field \cite{Heisenberg:1936nmg,Schwinger:1951nm}. The spectrum of produced particles is well studied for a time dependent electric field using a wide range of techniques \cite{Brezin:1970xf,Narozhnyi:1970uv,Marinov:1977gq,Kluger:1998bm,Gavrilov:1996pz,Kim:2000un,Ringwald:2001ib,Dunne:2008kc,Hebenstreit:2010vz,Gelis:2015kya}.
Quantum interference plays a key role in determining the particle production spectrum, and semiclassical intuition can be used for quantum control to design pulses with desired spectral characteristics \cite{Schutzhold:2008pz,Dumlu:2010ua,Akkermans:2011yn,Dumlu:2011rr}.

Moreover, the resulting quantum state is known to be entangled \cite{Ebadi:2014ufa,Florio:2021xvj,Nishida:2021qta}, and the entropy of entanglement between the particles and antiparticles is equal to the Gibbs entropy of the produced system \cite{Florio:2021xvj}.
In this work, we show that the effect of quantum interference is to \textit{decrease} the entanglement entropy.

The underlying physical phenomenon should appear in a number of quantum two-level systems, including
ionization of atoms and molecules \cite{Keldysh:1965ojf,2005PhRvL..95d0401L,Krausz:2009zz},
time-dependent tunneling \cite{Keski-Vakkuri:1996lbi}, Landau-Zener effect \cite{zueco2008landau,oka2009nonequilibrium,Shevchenko:2010ms}, driven atomic systems \cite{li2010carrier,jha2011experimental},
chemical reactions \cite{miller1968semiclassical,saha2011tunneling}, Hawking radiation \cite{Brout:1995rd,Parikh:1999mf,Volovik:2022cqk}, cosmological particle production \cite{Parker:1968mv}, heavy ion collisions \cite{Greiner1985-qt,Kharzeev:2005iz,Kharzeev:2006zm,Blaschke:2017igl},
shot noise in tunnel junctions \cite{Klich:2008un,https://doi.org/10.48550/arxiv.1611.06738}, and the
dynamical Casimir effect \cite{jaekel1997movement,dodonov2010current}. The suppression of entanglement entropy has potential practical application in different domains, including quantum-enhanced classical communication \cite{doi:10.1116/5.0036959}.  We treat both bosons and fermions as we envisage applications to quantum detectors with both bosonic and fermionic modes. We show that the leading suppression effect is the same.

\vskip0.3cm

\textit{Pair creation, entanglement entropy and multiple pulses.} We consider the phenomenon of pair creation in a background electric field $\vec{E}(t)=(0,0,E(t))$, with $\vec{A}(t)=(0,0,A(t))$ the associated gauge potential and $E(t)=-\dot{A}(t)$. Neglecting  backreaction on the field, one can  solve  the Klein-Gordon (respectively Dirac) equation for bosonic ($b$) (respectively fermionic ($f$)) fields. A natural  formalism is that of Bogoliubov transformations: see, e.g. \cite{Kluger:1998bm,Dumlu:2011rr}, and references therein.

Given some initial creation and anhihilation operators $a_k^{b/f},b_{-k}^{b/f\ \dagger}$, the Bogoliubov transformation coefficients $\alpha_k^{b/f}(t), \beta_k^{b/f}(t)$ relate them to the time dependent basis $\tilde a_k^{b/f}(t),\tilde b_{-k}^{b/f\ \dagger}(t)$
\begin{align}
  \begin{pmatrix}
    \tilde a_k^{b/f}(t) \\
    \tilde b_{-k}^{b/f\ \dagger}(t)
  \end{pmatrix}
  =
  \begin{pmatrix}
    \alpha_k^{b/f}(t) & \pm \beta_k^{b/f*}(t)\\
    \beta_k^{b/f}(t) &  \alpha_k^{b/f*}(t)
    \end{pmatrix}
  \begin{pmatrix}
  a_k^{b/f} \\
       b_{-k}^{b/f\ \dagger}
    \end{pmatrix} \ ,
\end{align}
with the $+/-$ sign for bosons/fermions. $\left| \beta_k^{b/f}(t)  \right|^2$ is the density of produced particles with momentum $k$. The bosonic/fermionic statistics are encoded in the constraint: $\left|\alpha_k^{b/f}(t)\right|^2 \mp \left| \beta_k^{b/f}(t)  \right|^2=1$. In particular, the $+$ sign in the fermionic case enforces the Pauli exclusion principle; the number of fermions per mode cannot exceed one.

The time evolution of the system can directly be rewritten in terms of the Bogoliubov coefficients, whose time evolution is that of a two-level system \cite{Marinov:1977gq,Kluger:1998bm}. Extracting suitable phases, $c_{\alpha, k}^{b/f}=e^{-i\int^t \dd \tau \en(\tau)}\alpha_k^{b/f}$, $c_{\beta, k}^{b/f}=e^{i\int^t \dd \tau \en(\tau)}\beta^{b/f}$, the evolution equations become \cite{Dumlu:2011rr,Akkermans:2011yn}
\begin{align}
  \frac{\dd}{\dd t} \begin{pmatrix}
    c_{\alpha, k}^{b/f} \\
    c_{\beta, k}^{b/f}
\end{pmatrix} &=
\begin{pmatrix}
  -i \en(t) & \Omega^{b/f}(t) \\
  \pm \Omega^{b/f}(t) & i\en(t)
\end{pmatrix}
\begin{pmatrix}
  c_{\alpha, k}^{b/f} \\
  c_{\beta, k}^{b/f}
\end{pmatrix} \ .
\label{eq:twolevelsys}
\end{align}
Here $\en(t)=\sqrt{m^2 +|\vec k_\perp|^2 + (k_{\parallel}-A(t))^2}$ is the dispersion relation
with $\vec k_\perp$ the momentum perpendicular to $\vec E(t)$ and $k_\parallel$ the longitudinal component. The frequencies  $\Omega^{b/f}(t)$ are read from the Klein-Gordon and Dirac equations and differ for bosons $\Omega^b(t)=E(t)(k_\parallel - A(t)) / (2 \en(t)^2)$, and  fermions $\Omega^f(t)=E(t)\sqrt{m^2+k_\perp} / (2 \en(t)^2)$. For simplicity, we set the transverse momentum to zero for the rest of this work: $k_\perp = 0, k_\parallel = k$.

The presence of the electric field creates a distinction between particles whose momenta is aligned with the electric field ("left movers") and particles whose momenta is antialigned with the electric field ("right movers"). An informative characterization of the entanglement present in the produced quantum state is obtained by computing the entanglement entropy between left and right movers \cite{Ebadi:2014ufa,Florio:2021xvj}

\begin{align}
    S&=-\int \frac{\dd k }{2\pi} \left [  \left|\alpha_k\right|^2 \log\left(\left|\alpha_k\right|^2 \right)+ \left|\beta_k\right|^2\log\left(\left|\beta_k\right|^2 \right)\right ]
  \label{eq:Sdecomposition}
\end{align}
We will use the notation $S_\alpha$ and $S_\beta$ for the first and second terms in (\ref{eq:Sdecomposition}).
Note that this left-right entanglement entropy has been found equal to the statistical Gibbs entropy of the produced pairs \cite{Florio:2021xvj}.

\begin{figure}
  \centering
  \includegraphics{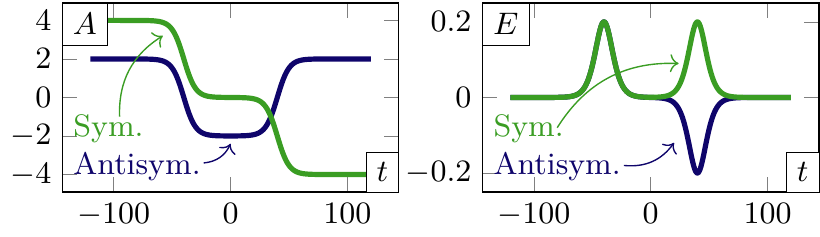}
  \caption{(Anti)-symmetric configurations for two pulses, $m=1, E=0.2, \tau=10, T=80$ (dimensionful quantities are expressed in units of $m$). \textbf{Left:} Vector potential. \textbf{Right:} Electric field.}
  \label{fig:fields}
\end{figure}

In this work, we focus on specific time sequences of pulses. We contrast symmetric configurations, where all the pulses have an electric field with the same sign, with antisymmetric configurations, where pulses have electric fields of alternating signs. Our analysis applies to very general temporal shapes of each individual pulse, but for definiteness we choose sequences of Sauter pulses, the pulse shape analyzed in \cite{Florio:2021xvj}. The basic physics can be seen in the 2-pulse configurations:  $A^{A}(t) = E\tau\left(1+\tanh\left(\frac{1}{\tau}\left(t-\frac{T}{2}\right)\right)-\tanh\left(\frac{1}{\tau}\left(t+\frac{T}{2}\right)\right)\right)$, $A^S(t)=-E\tau\left(\tanh\left(\frac{1}{\tau}\left(t-\frac{T}{2}\right)\right)+\tanh\left(\frac{1}{\tau}\left(t+\frac{T}{2}\right)\right)\right)$, $E^{A/S}(t)=-\dot{A}^{A/S}(t)$.
These 2-pulse configurations are illustrated in Fig.~\ref{fig:fields}.
The antisymmetric configuration, with an alternating electric field, is in blue, and the symmetric one is in green.
$E$ is the electric field amplitude, $\tau$ the duration of a single pulse and $T$ the separation between two consecutive pulses.
The corresponding $N$-pulse configurations are given in the Supplementary Material: see Eqs (\ref{eq:vec})-(\ref{eq:fld}).

For notational convenience, we further define $n^{b/f,A/S}_N(k)\equiv\left |\beta^{b/f,A/S}_k\right|^2$, which is the bosonic/fermionic number density per mode of particles created by the antisymmetric (A) or symmetric (S) configuration of $N$ pulses.  We consider the semiclassical limit $E\ll m^2, E\tau \gg m$, in which pair creation occurs by non-perturbative tunneling from the Dirac sea, and we study the quantum interference effects by focussing on well-separated pulses: $\tau\ll T$. In this case, the particle number is small and can be computed in the semiclassical approximation \cite{Brezin:1970xf,Marinov:1977gq,Dumlu:2010ua,Akkermans:2011yn,Dumlu:2011rr}. The solutions localize around complex "turning points" $t_p$, defined by $\mathcal{E}(t_p)=0$, where the phase $\int^t {\mathcal E}$ is approximately stationary. Thus, for a single pulse \cite{Brezin:1970xf,Narozhnyi:1970uv,Marinov:1977gq}
\begin{align}
  n_1(k) \approx \exp\left(-2\left|\int_{t_0}^{t_0^*}\dd t\mathcal{E}(t) \right|\right) \ ,
\end{align}
with $t_0$ the turning point closest to the real axis.
For example, the case $A(t)= -E\tau \tanh\left(\frac{1}{\tau}\left(t-T\right)\right)$ gives $t_0|_T = T + \tau \mathrm{arctanh}\left(\frac{-(k- i m)}{E\tau}\right)$.

\begin{figure}
  \centering
  \includegraphics{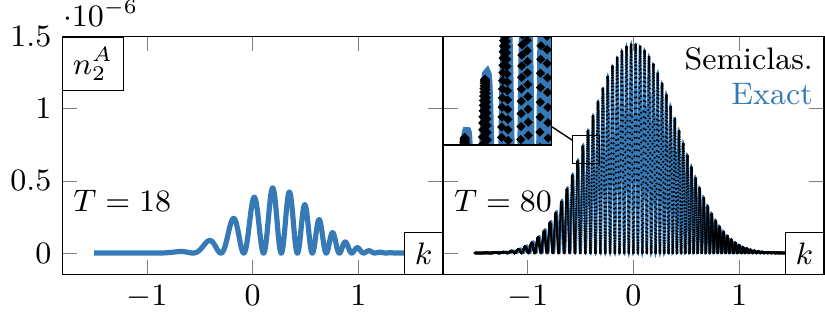}
  \caption{Spectra of produced fermionic particles for two pulses in the antisymmetric configuration. \textbf{Left:} Overlapping pulses. \textbf{Right:} Well-separated pulses. The interference pattern is clear and the semiclassical expression works well for separated pulses.}
  \label{fig:antispectra}
\end{figure}

\begin{figure}
  \centering
  \includegraphics{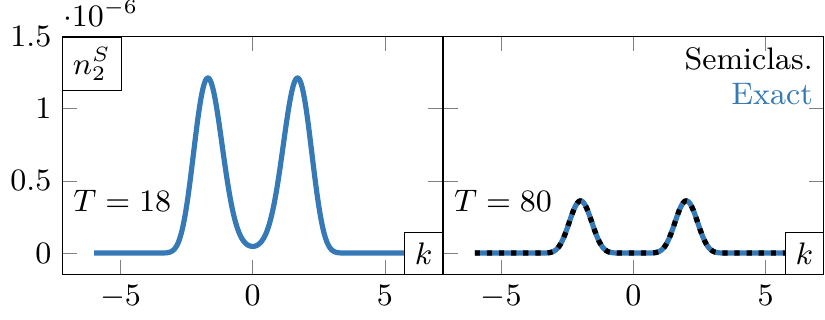}
  \caption{Spectra of produced fermionic particles for two pulses in the symmetric configuration. \textbf{Left:} Overlapping pulses. \textbf{Right:} Well-separated pulses. The semiclassical approximation works in the latter case and is shown on top of the exact solution.}
  \label{fig:symspectra}
\end{figure}

For {\it sequences} of multiple pulses, interference effects can arise, which can be understood semiclassically \cite{Akkermans:2011yn}. The situation is very different for the antisymmetric (A) or symmetric (S) pulse sequences. Numerically computed particle momentum spectra for the case $N=2$ are shown in Fig.~\ref{fig:antispectra} and Fig.~\ref{fig:symspectra}. For the antisymmetric configuration, interferences are strong and the spectrum is highly oscillatory, whereas for the symmetric configuration, there is no interference for well-separated pulses. The bosonic case is qualitatively similar, up to the expected Maslov phase difference for the antisymmetric configuration.

 These particle spectra are well described by the semiclassical approximation,
in which the particle number is obtained as the modulus squared of a sum of amplitude contributions
$A_p$ from the different turning points $n(k)\approx \left|\sum_p A_p(k)\right|^2$. In the symmetric case and for large enough separation $T$, the dominant turning points are independent, for any given $k$. The resulting spectrum is therefore effectively a sum of  single pulse spectra
\begin{align}
n_N^{b/f,S}(k) \approx \sum_{l=1}^N n_1|_{(l)}(k) \ .
\label{eq:symsemi}
\end{align}

The situation is very different in the antisymmetric case.  For any given $k$, $N$ turning points contribute with equal strength but different phases, leading to coherent interference. The resulting particle spectrum is well approximated by a Fabry-Perot-like form \cite{Akkermans:2011yn}
\begin{align}
  n_N^{f,A}(k)&=\begin{cases}
  n_1(k) \frac{\sin^2\left(N\phi_k\right)}{\cos^2\left(\phi_k\right)} \ , \ N \text{ even}  \\
  n_1(k) \frac{\cos^2\left(N\phi_k\right)}{\cos^2\left(\phi_k\right)} \ , \ N \text{ odd}
\end{cases}
\label{eq:antisemi-f}
\\ 
n_N^{b,A}(k)&=
  n_1(k) \frac{\sin^2\left(N\phi_k\right)}{\sin^2\left(\phi_k\right)} \ .
  \label{eq:antisemi-b}
\end{align}
In (\ref{eq:antisemi-f})-(\ref{eq:antisemi-b}), $\phi_k$ is the semiclassical phase difference between two turning point pairs $t_{\pm}$
\begin{align}
  \phi_k=\int_{Re(t_-)}^{Re(t_+)}\dd t &\sqrt{m^2 + (k-A^A(t))^2} \ . \label{eq:phik}
\end{align}
The constant phase difference between the bosonic and fermionic case results from the behavior of the effective potential around the turning point. It is linear in the fermionic case and quadratic in the bosonic case, leading to different Maslov indices \cite{Froman:1970toy,Meyer:1976uj,Dumlu:2010ua,Akkermans:2011yn}. Thus Eqs.~(\ref{eq:antisemi-f})-(\ref{eq:antisemi-b}) can be rewritten as
\begin{align}
    n_N^{b/f,A}(k) = \left(N + 2\sum_{n=1}^{N-1}(\pm 1)^n (N - n)\cos(2 n \phi_k)\right ) n_1(k)
    \label{eq:antisemibis}
\end{align}
with $\phi_k$ defined in Eq.~\eqref{eq:phik}. The $+$ sign for boson and $-$ sign for fermions arise from the constant phase in Eq.~\eqref{eq:phik}.
\\

\begin{figure}
  \centering
  \includegraphics{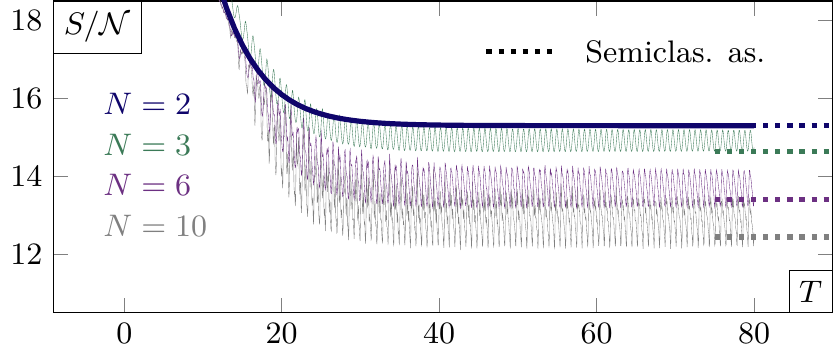}
  \caption{Entropy to the particle number ratio in the fermionic case, $m=1, E=0.2, \tau=10$ (dimensionful quantities are expressed in units of $m$). Antisymmetric configuration of $N=2,3,6,10$ pulses. The asymptotes are also correctly predicted by the semiclassical expression \eqref{eq:suppfactorsemi}. The presence of increasingly stronger oscillations as a function of $N$ is an indication of interferences between multiple turning points, an effect not included in the semiclassical spectrum \eqref{eq:antisemi-f}. The bosonic case leads to qualitatively similar results and is not shown.}
  \label{fig:ratioN}
\end{figure}

\textit{Entropy suppression.} We now turn to the effect of quantum interference on the entanglement/Gibbs entropy (\ref{eq:Sdecomposition}) of the produced particles. Recall that $|\beta_k|^2=n(k)$, and $|\alpha_k|^2=1\pm |\beta_k|^2$ for bosons/fermions,  so the entropy can be computed directly from the particle spectra.
We are most interested in the case of well-separated pulses where the quantum interference effects are strong (recall Figs. \ref{fig:antispectra} and \ref{fig:symspectra}).
\begin{figure}
  \centering
  \includegraphics{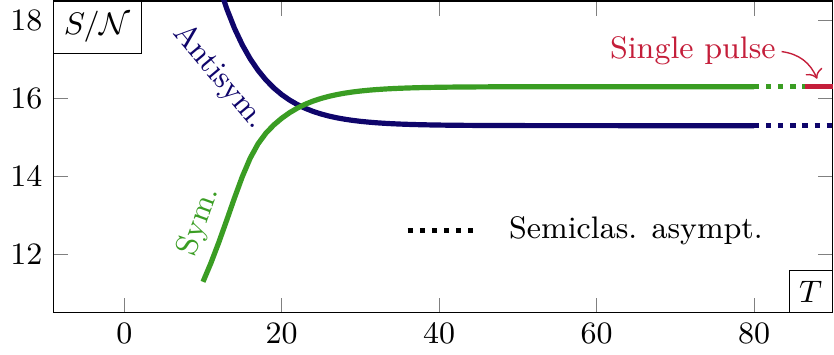}
  \caption{Entropy to particle number ratio in the fermionic case, $m=1, E=0.2, \tau=10$ (dimensionful quantities are expressed in units of $m$), for antisymmetric and symmetric configurations of two pulses. The symmetric configuration asymptotes to the single pulse ratio. The presence of interference reduces the entropy per particle and the antisymmetric configuration asymptotes to a lower ratio, correctly predicted by the semiclassical expression \eqref{eq:suppfactortwo}. The bosonic case leads to qualitatively similar results and is not shown.}
  \label{fig:ratiotwo}
\end{figure}
The numerical results for the ratio of entropy to particle number are shown in Fig.~\ref{fig:ratiotwo} for the symmetric and antisymmetric configurations of two pulses, as a function of the temporal separation $T$. The solid lines show the result of solving numerically the two-level system equations \eqref{eq:twolevelsys} using \texttt{DifferentialEquations.jl}  \cite{rackauckas2017differentialequations}. We see that for well-separated pulses, the entropy of the antisymmetric configuration asymptotes to a lower value than the symmetric one. In other words, \textit{the presence of quantum interference results in a decrease of the entanglement/Gibbs entropy}.
Corresponding results are shown in Fig.~\ref{fig:ratioN} for the antisymmetric configuration with further sequences of pulses.

The details of these entropy suppression effects can be explained using semiclassical arguments. We start by discussing the total number of produced particles: $\mathcal{N}_N^{b/f,A/S}:=\langle n^{b/f,A/S}_N\rangle = \int \frac{\dd k}{2\pi} n^{b/f,A/S}_N(k)$. In the limit of large time separation, we find that
 \begin{align}
  \mathcal{N}^{b/f,A/S} \approx N \, \mathcal{N}_1^{b/f}\ . \label{eq:asNum}
\end{align}
This is easily seen in the symmetric case (recall \eqref{eq:symsemi} and Fig.~\ref{fig:symspectra}), but is more generally true due to the fact that
the semiclassical phase becomes a rapidly varying function of $k$. Therefore, all integrals of the type $\int \dd k f(k) e^{i\phi_k}$ are asymptotically exponentially suppressed. Physically, the modes are re-distributed due to quantum interference between separated pulses, but the total number simply scales with $N$ \cite{Akkermans:2011yn}.

The entanglement entropy \eqref{eq:Sdecomposition} has a more interesting dependence on the interference.
The symmetric configuration is the simplest, as there is no quantum interference at large separation, so all powers of the spectral density just scale with the number of pulses $N$. Therefore
\begin{eqnarray}
S_N^{b/f, S}\approx N \, S_1^{b/f} \quad .
\label{eq:S-S}
\end{eqnarray}
Combined with \eqref{eq:asNum}, this explains the fact that in Fig.~\ref{fig:ratiotwo} we see that for the symmetric configuration the entropy to particle number ratio tends to that of a single pulse.

For the antisymmetric configuration, we focus first on  $N=2$ (as in Fig.~\ref{fig:ratiotwo}) and then give the generalization to $N$ alternating sign pulses. For the $N=2$ antisymmetric pulse case, we can write the particle spectra (\ref{eq:antisemi-f})-(\ref{eq:antisemi-b}) as $n_2^{f, A}(k)=4 \, n_1(k)\, \sin^2(\phi_k)$ and $n_2^{b, A}(k)=4 \, n_1(k)\, \cos^2(\phi_k)$. Integrating over $k$, we find that expectations of powers of the density scale in a universal way in the semiclassical limit:
 \begin{eqnarray}
   \left\langle \left(n_2^{b/f,A}\right)^p\right\rangle  &\sim \frac{(2p)!}{(p!)^2} \left \langle \left(n_1^{b/f}\right)^p \right\rangle\ .\label{eq:moment2pulses}
 \end{eqnarray}
Furthermore, the contributions from $S_{N,\alpha}^{b/f, A}$, the first term in \eqref{eq:Sdecomposition}, are subdominant in the semiclassical limit. Therefore, we find that
 \begin{eqnarray}
   S_{2}^{b/f, A} \approx S_{\beta,2}^{b/f, A}  
   &=&\left . \int \frac{\dd k}{2\pi} \frac{\dd}{\dd p} \left(n_2^{b/f,A}(k)\right)^p\right|_{p=1} \label{eq:ddpentropy} \\
   &\approx & 2 S_1^{b/f} - 2 \mathcal{N}_1^{b/f} \ . \label{eq:suppfactortwo}
 \end{eqnarray}
Combined with \eqref{eq:asNum}, this explains the numerical observation in Fig.~\ref{fig:ratiotwo} that for the antisymmetric 2-pulse configuration the entropy to particle number ratio is approximately reduced by 1 due to quantum interference.

An analogous argument (see the Supplementary Material)
for the $N$-pulse antisymmetric configuration yields a universal leading suppression
\begin{align}
  S^{b/f,A}_N \approx N\,  S_1^{b/f} -2\, \mathcal{N}^{b/f}_1 (1-N+N\, H_{N-1}) \ , \label{eq:suppfactorsemi}
\end{align}
Here $H_{N}=\sum_{n=1}^{N}\frac{1}{n}$ is the $N^{th}$ harmonic number. This explains the behavior in the exact numerical results  in Fig.~\ref{fig:ratioN}. The increasingly large oscillations are due to the interference beyond neighboring turning points, not accounted for in the semiclassical spectrum \eqref{eq:antisemi-f}.

 In particular, notice that for the far-separated antisymmetric pulse configuration the entanglement entropy \eqref{eq:suppfactorsemi} is always \textit{less} than for the symmetric configuration \eqref{eq:S-S}: quantum interference leads to entropy suppression.

\textit{Discussion.} We have shown that the effect of quantum interference on the process of Schwinger pair creation is to \textit{reduce} the entropy of the produced quantum state. This remarkable result directly generalizes to other two-level systems, as no specific assumptions beyond the validity of the semiclassical approximation were made about the type of the process or shape of the incoming signal.

An interesting theoretical avenue is to study this effect in dynamical systems. In particular, linking quantum interference to entanglement spreading may provide valuable insights into the study of information scrambling, see Ref.~\cite{Swingle2018QuantumIS} for a review. A similarly exciting prospect is to study the effect of backreaction using quantum computations in $1+1$ directions.

Another more direct and concrete outlook is to utilize this effect in the context of information transmission. Classical information enhanced by quantum detection is an active field of research, see \cite{doi:10.1116/5.0036959} for a review.  Considering the produced quantum excitations as the state of a quantum receiver, our result suggests that classical signals with the same entropy generate
quantum states with different entropies. Note that the leading effect is independent of the quantum statistics of the produced particles. This can be recast as the fact that the mutual information $\mathcal{I}^{A/S}_N$ \cite{Shannon:1948zz,nielsen2010quantum,akkermans2009transmission} of the antisymmetric
 configuration is larger than the symmetric configuration, $\mathcal{I}^{A}_N - \mathcal{I}^{S}_N\sim 2 \mathcal{N}^1 (1-N+NH_{N-1})>0$ by Eq.~\eqref{eq:suppfactorsemi}, opening up the possibility of improved efficiencies.   Similar prospects are offered by optimized pulse shapes \cite{levitov1996electron}.
The effect on the entropy of (quasi)-periodic driving \cite{doi:10.1080/00018738200101358,gabelli2013shaping,Verdeny_2016,dumitrescu2018logarithmically,dumitrescu2022dynamical}, known to produce spectra (sometimes exponentially) more localized than our antisymmetric configuration, as well as the link between entropy suppression and localization, are equally interesting. We leave these for future work.

\textit{Acknowledgements:}
This work was supported by the U.S. Department of Energy, Office of Science, High Energy Physics Program under Award DE-SC0010339 (GD), and Nuclear Physics Program under Awards DE-SC0012704 (AF) and
DE-FG88ER41450 and DE-SC0012704 (DK). This material is based upon work supported by the U.S. Department of Energy, Office of Science, National Quantum Information Science Research Centers, Co-design Center for Quantum Advantage (C2QA) under Contract No. DE-SC0012704.


\appendix

\clearpage
\onecolumngrid

\section*{Supplementary Material}
\label{app:npulses}

The analysis in the main text for 2 antisymmetric (A) or symmetric (S) pulses generalizes to $N$ pulses as follows. We choose alternating and non-alternating sequences of Sauter electric field pulses, with vector potentials and electric fields:
\begin{align}
  A^{A/S}(t) &= \begin{cases} E \tau \left\{\eta^{A/S}+\sum_{n=1}^{N/2}(-1)^n \left[\tanh\left(\frac{1}{\tau}\left(t+\left(n-\frac{1}{2}\right)T\right)\right)\mp \tanh\left(\frac{1}{\tau}\left(t-\left(n-\frac{1}{2}\right)T\right)\right)\right]\right\}\\
   E \tau \left(\sum_{n=-(N-1) / 2}^{(N-1) / 2}(\mp 1)^n\tanh\left(\frac{1}{\tau}\left(t+nT\right)\right) \right)
  \end{cases}\label{eq:vec}\\
  E^{A/S}(t) &=\begin{cases}- E \sum_{n=1}^{N/2}(-1)^n \left[\cosh^{-2}\left(\frac{1}{\tau}\left(t+\left(n-\frac{1}{2}\right)T\right)\right)\mp \cosh^{-2}\left(\frac{1}{\tau}\left(t-\left(n-\frac{1}{2}\right)T\right)\right)\right]\\
  - E  \left(\sum_{n=-(N-1) / 2}^{(N-1) / 2}(\mp 1)^n\cosh^{-2}\left(\frac{1}{\tau}\left(t+nT\right)\right) \right) \ .
\end{cases}\label{eq:fld}
\end{align}

The constant $\eta^{A}=(-1)^{N/2+1},\ \eta^{S}=0$ is a shift introduced in the vector potential solely to center the momentum spectra around the origin. It has no physical consequence in the situation at hand.

To explain in simple physical terms the phenomenon of entropy suppression for the antisymmetric pulse configuration, $E^A(t)$, we first consider an analytic model that captures the essential physics. We analyze the fermionic case for $N$ even, and the other cases follow similarly. We consider a single pulse that produces a localized particle momentum spectrum $n_1^f(k)$ with associated  particle number $\mathcal N_1^f$:
\begin{eqnarray}
{\mathcal N}_1^f\equiv \langle n_1^f\rangle =\int_{-\infty}^\infty \frac{dk}{2\pi} \, n_1^f(k)
\label{eq:n1f}
\end{eqnarray}
Consider for example $n_1^f(k)=2 {\mathcal N}_1^f/(k^2+1)$, or $n_1^f(k)=\sqrt{4\pi} {\mathcal N}_1^f\, e^{-k^2}$.
The associated entanglement entropy is
\begin{eqnarray}
S_1^f = -\langle n_1^f\log(n_1^f)\rangle
 -\langle (1-n_1^f)\log(1-n_1^f)\rangle
 \label{eq:s1f}
\end{eqnarray}
For the 2-pulse antisymmetric configuration, we model the interference for well-separated pulses as (recall Fig.~\ref{fig:antispectra}):
\begin{eqnarray}
    n_2^{f, A}(k)= 4 \sin^2(T k) \, n_1^{f}(k)
    \label{eq:n1fa}
\end{eqnarray}
Then, by a stationary phase argument, we see that at large pulse separation $T$ the powers of $(n_2^{f, A})$ behave as:
\begin{eqnarray}
\langle (n_2^{f, A})^p\rangle \equiv \int_{-\infty}^\infty \frac{dk}{2\pi}  (n_2^{f, A}(k))^p \to
\frac{(2p)!}{(p!)^2} \langle n_{1}^p\rangle
\label{eq:as-moments}
\end{eqnarray}
We deduce that:
\begin{eqnarray}
\langle n_2^{f, A}\rangle &\to& 2 \langle n_{1}\rangle
\label{eq:identities1}\\
\langle n_2^{f, A}\log(n_2^{f, A})\rangle &\to&  2\langle n_{1}^{f}\, \log(n_{1}^{f})\rangle + 2\langle n_{1}^f\rangle
\label{eq:identities2}
\end{eqnarray}
where we have used the fact that $n\log(n) =\left.\frac{d}{dp} n^p\right|_{p=1}$.
Furthermore, expanding the logarithm, we find
\begin{eqnarray}
\langle (1-n_2^{f, A})\log(1-n_2^{f, A})\rangle &=&
-\langle n_2^{f, A}\rangle
+\sum_{p=2}^\infty \frac{\langle (n_2^{f, A})^p\rangle}{p(p-1)} 
\nonumber\\
&\to & -2\langle n_{1}^f\rangle +\sum_{p=2}^\infty \frac{(2p)!}{(p!)^2}
\frac{\langle (n_{1}^f)^p\rangle}{p(p-1)}
\nonumber\\
&=&2\langle (1-n_{1}^f)\log(1-n_{1}^f)\rangle +\sum_{p=2}^\infty \left(\frac{(2p)!}{(p!)^2}-2\right)\frac{\langle (n_{1}^f)^p\rangle}{p(p-1)}
\nonumber
\end{eqnarray}
Therefore, the entanglement entropy behaves as
\begin{eqnarray}
S_2^{f, A} \to 2 S_{1}^f -2\langle n_{1}^f\rangle -\sum_{p=2}^\infty \left(\frac{(2p)!}{(p!)^2}-2\right)\frac{\langle (n_{1}^f)^p\rangle}{p(p-1)}
\label{eq:sfa2}
\end{eqnarray}
Thus, the entropy/number ratio for the antisymmetric configuration is {\it reduced} compared to the one-pulse case. Since $\langle (n_{1}^f)^p\rangle$ is small, the first correction term, $-2\langle n_{1}^f\rangle$, dominates. This explains the numerical result in Fig.~\ref{fig:ratiotwo} for the large $T$ entropy/number ratio:
\begin{eqnarray}
 \frac{S_2^{f, A}}{\langle n_2^{f, A}\rangle} \approx \frac{S_1^f}{\langle n_{1}^f\rangle} -1 \ .
 \label{eq:s2fa-dominant}
 \end{eqnarray}
By contrast, for the well separated symmetric 2-pulse configuration (recall Fig.~\ref{fig:symspectra}) we have $\langle (n_2^{f, A})^p\rangle \to 2\langle (n_1^f)^p\rangle$, so that
\begin{eqnarray}
\frac{S_2^{f, S}}{\langle n_2^{f, S}\rangle} \approx \frac{S_1^f}{\langle n_{1}^f\rangle}
 \label{eq:s2fs-dominant}
\end{eqnarray}
as observed numerically in Fig.~\ref{fig:ratiotwo}. For bosonic fields the argument is similar, with some sign changes, but the leading behavior of the entropy/number ratio is the same as in \eqref{eq:s2fa-dominant}-\eqref{eq:s2fs-dominant}.

For the $N$-pulse antisymmetric and symmetric configurations we can again use a stationary phase argument. We model the interference with the Fabry-Perot form:
\begin{eqnarray}
n_N^{f, A}(k)= \frac{\sin^2(N T k)}{\cos^2(k T)} \, n_1^f(k)
\label{eq:nnfa}
\end{eqnarray}
Then the moment formula (\ref{eq:as-moments}) becomes:
\begin{eqnarray}
\langle (n_N^{f, A})^p\rangle  &\to& C_{p, N} \,  \langle (n_{1}^f)^p\rangle
\label{eq:n-a-moments}
\end{eqnarray}
where the coefficient $(2p)!/(p!)^2$ in (\ref{eq:as-moments}) generalizes to
\begin{equation}
C_{p, N}=\hskip -5pt \sum_{j=0}^{\left[\frac{p(N-1)+1}{N}\right]}   \hskip -5pt  \frac{(-1)^j(2p) \Gamma(N(p-j)+p)}{j!\Gamma(N(p-j)+1-p) \Gamma(2p-j+1)}
\label{eq:as-moments-n}
\end{equation}
The identities (\ref{eq:identities1})-(\ref{eq:identities2}) generalize to:
\begin{eqnarray}
\langle n_N^{f, A}\rangle &\to& N \langle n_{1}^f\rangle \\
\langle n_N^{f, A}\log(n_N^{f, A})\rangle &\to&  N\langle n_{1}^f\, \log(n_{1}^f)\rangle + 2(1-N+N H_{N-1})\langle n_{1}^f\rangle
\label{eq:identities-n}
\end{eqnarray}
where $H_N$ is the harmonic number. Therefore, for the $N$-pulse alternating sign configuration the entanglement entropy behaves as
\begin{eqnarray}
S_N^{f, A} &\to& N S_{1}^f -2(1-N+N H_{N-1})\langle n_{1}^f\rangle -\sum_{p=2}^\infty \frac{\left(C_{p, N}-N\right)}{p(p-1)}\langle (n_{1}^f)^p\rangle
\label{eq:snfa}
\end{eqnarray}
The correction terms are all negative, and for small number densities, as is the case in the semiclassical limit, the dominant effect in (\ref{eq:snfa}) is from the first correction term, proportional to $\langle n_{1}^f\rangle$. This explains the key physics of the numerical behavior seen in Fig.~\ref{fig:ratioN}. The corresponding result for bosons is similar:
\begin{eqnarray}
S_N^{b, A} &\to& N S_{1}^b -2(1-N+N H_{N-1})\langle n_{1}^b\rangle -\sum_{p=2}^\infty (-1)^p \frac{\left(C_{p, N}-N\right)}{p(p-1)}\langle (n_{1}^b)^p\rangle
\label{eq:snba}
\end{eqnarray}

To connect this model interference analysis with the full semiclassical analysis, we replace the interference effect on the momentum spectrum in (\ref{eq:nnfa}) by
\begin{eqnarray}
n_N^{f, A}(k)= \frac{\sin^2(N \phi_k)}{\cos^2(\phi_k)} \, n_1^f(k)
\label{eq:nnfa-semi}
\end{eqnarray}
where $\phi_k$ is the interference phase in (\ref{eq:phik}).
We rescale $\phi_k$ to make it $O(1)$, and define $\tilde\phi_k =\frac{\phi_k}{T}$. In these terms, we are dealing with integrals of the form $I=\int \dd k \, n_1(k) e^{iT\tilde\phi_k}$. For a smooth spectrum $n_1(k)$, we can again use a stationary phase approximation, so that an isolated stationary point $k^*$ yields
$I\approx n_1(k^*)e^{i\phi_{k^*}} + O(\frac{1}{T})$ \cite{10.2307/j.ctt1bpmb3s.13}. Exponential suppression occurs when $k^*$ is parametrically separated from $k_0$, with $k_0$ the momentum the spectrum is centered around.

This is indeed what happens for our pulse profiles \eqref{eq:vec}-\eqref{eq:fld}. For large $T$, the real part of the turning points is well approximated by $Re(t_\pm)=\pm\frac{T}{2}$. Neglecting the $k$ dependence in the turning points, we have $\phi_k'\approx \int_{-T/2}^{T/2} \frac{k-A(t)}{\sqrt{m^2+(k-A(t))^2}}$. Noting that $\max(A(t))=E\tau$, $\min(A(t))=-E\tau$, we have   $\int_{-T/2}^{T/2} (k-A(t)) > \phi_k' > \frac{1}{\sqrt{m^2+(k\pm E\tau)^2}} \int_{-T/2}^{T/2} (k-A(t))$, with the $\pm$ sign depending on $k$ being positive or negative.
As a result, $k_*\approx \int_{-T/2}^{T/2}\dd t A(t)=-E \tau + O(1/E\tau)$. Our spectrum is centered around $k_0=0$ and we work in the limit $E\tau\gg m$, leading to $n_1(k_*)\ll n_1(k_0)$.
\begin{figure}
  \centering
  \includegraphics{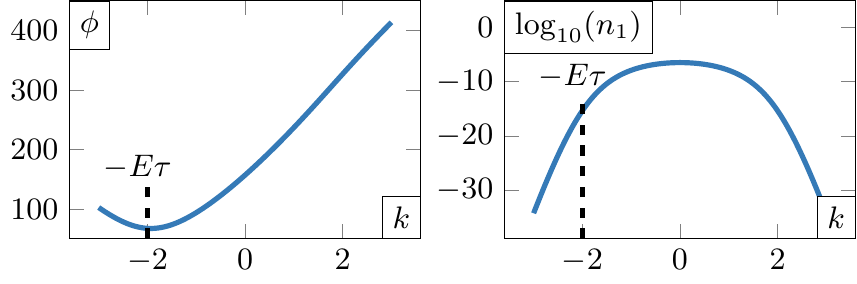}
  \caption{Semiclassical phase and single pulse spectrum for $m=1, E=0.2, \tau=10, T=80$ (dimensionful quantities are expressed in units of $m$). \textbf{Left:} Semiclassical phase. The minimum is well approximated by $k_*=E\tau$. \textbf{Right:} Logarithm of the single pulse spectrum. We have $n_1(E\tau)\approx 5\cdot10^{-16}$ and $n_1(0)\approx 4\cdot10^{-7}$, which shows that the phase integrals can be safely neglected. }
  \label{fig:semiclassicPhase}
\end{figure}
These results are confirmed in Fig.~\ref{fig:semiclassicPhase}. The minimum of the phase (left-hand side) is well approximated by $E\tau$ and the semiclassical single pulse spectrum is exponentially smaller at the minimum than at the center, $n_1(E\tau)\approx 5\cdot10^{-16}$ and $n_1(0)\approx 4\cdot10^{-7}$.

\textit{Alternative derivation:}
The suppression factor to $S_{\beta,N}$ for the antisymmetric pulse configurations can also be derived in an alternative fashion, inspecting directly the integral form of the entropy \eqref{eq:Sdecomposition}. Again, for simplicity, we focus on the fermionic even $N$ case; the final result is also valid for $N$ odd and bosonic particles.
Using the expressions \eqref{eq:antisemi-f}-\eqref{eq:antisemibis} in the main text, we can explicitly isolate the contribution from the single pulses and the contribution coming from interferences by rewriting $S_\beta$ as
\begin{align}
S^{b/f,A}_{\beta, N} &= N S_{\beta, 1} \\
&-N\int\dd k n_1(k) \log\left(\frac{\sin^2\left(N\phi_k\right)}{\cos^2\left(\phi_k\right)}\right)-2\int \dd kn_1(k)\sum_{n=1}^{N-1}(-1)^n(N-n)\cos(2 n \phi_k)\log(n_1(k))\notag\\
&-2\sum_{n=1}^{N-1} (- 1)^n(N-n)\Bigg[  \notag \int\dd kn_1(k) \cos(2 n \phi_k)\log\left(\sin^2\left(N\phi_k\right)\right)-\int\dd kn_1(k) \cos(2 n \phi_k)\log\left(\cos^2\left(\phi_k\right)\right)\Bigg] \\
&:= N S_{\beta, 1} + I_1 + I_2 + I_3 + I_4 \ .
\end{align}
The asymptotes of the interference integrals $I_1,I_2,I_3,I_4$ can be evaluated explicitly. Because of the fast oscillations, it is clear that $I_2\sim 0$. To compute the other asymptotes, we expand the logarithms in trigonometric functions. The only terms contributing to the asymptotes are the ones with no phase factor
\begin{align}
  I_1&=N\int\dd k n_1(k)\sum_{n=1}^\infty\frac{1}{n2^{2n}}\left[\left(e^{iN\phi_k}+ e^{-iN\phi_k}\right)^{2n}
    -(-1)^n\left(e^{i\phi_k}- e^{-i\phi_k}\right)^{2n}\right]\\
    &\sim N\mathcal{N}_1 \sum_{n=1}^\infty\frac{1}{n2^{2n}}\left[\frac{(2n)!}{(n!)^2} -\frac{(2n)!}{(n!)^2}\right] = 0 \ ,
\end{align}
where we used the binomial formula to extract the constant terms in the expansion.  The same logic applies to  $I_3$:
\begin{align}
  I_{3}=\sum_{n=1}^{N-1}\int\dd kn_1(k)(- 1)^n(N-n) \sum_{l=1}^{\infty} \frac{1}{2^{2l}l}\sum_{m=0}^{2l}{2l\choose m}\left(e^{i2\phi_k(n+Nm-Nl)}+e^{i2\phi_k(-n+Nm-Nl)}\right)\sim 0 \ ,
\end{align}
as $\pm n + Nm - Nl = 0$ has no solutions for $n\in [1\dots N-1]$. The only nonvanishing contribution is given by $I_4$:

\begin{align}
  I_{4}&=-\sum_{n=1}^{N-1}\int\dd k|\beta_k^{1}|^2(-1)^n(N-n)
  \sum_{l=1}^{\infty} \frac{(-1)^{l+m}}{2^{2l}l}\sum_{m=0}^{2l}{2l\choose m}\left(e^{i2\phi_k(n+m-l)}+e^{i2\phi_k(-n+m-l)}\right)\notag\\
  &\sim \mathcal{N}_1\sum_{n=1}^{N-1}(n-N) \sum_{l=1}^{\infty} \frac{1}{2^{2l}l}\left[{2l\choose l-n}+{2l\choose l+n}\right]
  \notag\\
  &=2\mathcal{N}_1\sum_{n=1}^{N-1}\frac{(n-N)}{n} = -2 \mathcal{N}_1 (1-N+NH_{N-1}) \ ,
\end{align}
using  $\sum_{l=1}^{\infty} \frac{1}{2^{2l}l}\left[{2l\choose l-n}+{2l\choose l+n}\right] =\frac{2}{n}$. It indeed leads to the same entropy suppression as in Eq.~\eqref{eq:identities-n}.

\end{document}